\documentclass{iaus}
\usepackage{graphicx}
\usepackage{natbib}
\usepackage{url}
\usepackage{hyperref}

\usepackage{multicol}

\def\apj{ApJ}%
\def\apjl{ApJ}%
\def\apjs{ApJS}%
\def\aap{A\&A}%
\def\mnras{MNRAS}%
\def\nar{New A Rev.}%
\def\hi{H{\sc I}}
\def\himf{H{\sc I}MF}

\title[MeerKAT's Deep \hi \ Field]{Looking At the Distant Universe with the MeerKAT Array (LADUMA)}
\author[B.W. Holwerda et al.]   
{B.W. Holwerda$^1$, S.--L. Blyth$^2$, A. J. Baker$^3$ \and the LADUMA team}

\affiliation{$^1$ European Space Agency (ESTEC),
Keplerlaan 1, 2200 AV Noordwijk, The Netherlands\\

email: {\tt \href{mailto:benne.holwerda@esa.int}{benne.holwerda@esa.int}} \\[\affilskip]
$^2$ ACGC, Department of Astronomy, University of Cape Town\\
$^3$ Rutgers University.\\
}

\pubyear{2011} 
\volume{284}  
\pagerange{1--12}
\jname{The Spectral Energy Distribution of Galaxies}
\editors{R.J. Tuffs \&  C.C.Popescu, eds.}

\begin{document}

\maketitle

\begin{abstract}
The MeerKAT (64 x 13.5m dish radio interferometer) is South Africa's precursor instrument for the Square Kilometre Array (SKA), exploring dish design, instrumentation, and 
the characteristics of a Karoo desert site and is projected to be on sky in 2016. One of two top-priority, Key Projects is a single deep field, integrating for 5000 hours total with the aim to detect neutral atomic hydrogen through its 21 cm line emission out to redshift unity and beyond.
This first truly deep HI survey will help constrain fueling models for galaxy assembly and evolution. It will measure the evolution of the cosmic neutral gas density and its distribution over galaxies over cosmic time, explore evolution of the gas in galaxies, measure the Tully-Fisher relation, measure OH maser counts, and address many more topics.
Here we present the observing strategy and envisaged science case for this unique deep field, which encompasses the Chandra Deep Field-South (and the footprints of GOODS, GEMS and several other surveys) to produce a singular legacy multi-wavelength data-set.


\end{abstract}

\firstsection 
\section{Introduction}

Over the past 8 Gyr, galaxies have evolved dramatically, with the number density of
luminous, blue, star-forming galaxies dropping by an order of magnitude \citep[e.g.,][]{Bell05}
and the number of luminous red galaxies increasing by a similar amount \citep{Driver98, Bell05}.  
%
Similarly since $z\sim1$, the star formation rate density in the Universe has plummeted by an order 
of magnitude \citep[e.g.,][]{Madau98, Hopkins08}, yet the density of neutral hydrogen (\hi), the fuel reservoir for star-formation, may have remained constant (Fig. \ref{f:Omhi}). Given that, locally, the overall \hi \ mass and star-formation rate appear to be linked for galaxies \citep{Kennicutt98}, this points to significant redistribution of the gas reservoir over this epoch.

To complement our view of high redshift galaxy populations (the stellar, ionised, and molecular gas), a key science goal of the Square Kilometre Array \citep[SKA][]{ska} is to explore the atomic hydrogen (\hi) out to $z\sim1$ and to explore how galaxies form and evolve. 
%
%
%
While SKA will study \hi \ in large numbers of galaxies at $z\sim1$, MeerKAT \citep{MEERKAT, meerkat2}, with its high sensitivity and wide instantaneous bandwidth, is the choice pathfinder instrument with which to begin the study of \hi \ in galaxies out to redshift unity. We therefore are undertaking the LADUMA\footnote{Literally, ÒIt thunders.Ó in the Zulu language: \url{http://www.ast.uct.ac.za/laduma/Home.html}} 
(Looking At the Distant Universe with the MeerKAT Array) survey for a 5000 hour observation of a single field
\citep{Holwerda10iska, Holwerda10vuvu, Holwerda11aas}.
This survey in combination with existing and planned complementary multi-wavelength observations will provide
the first complete picture of galaxy evolution from $0 < z \lesssim1.4$ and serve as a key benchmark for
future studies with the SKA.

\section{Observing Strategy}

\noindent \underline{Target Field} The Extended Chandra Deep Field-South was chosen for its wealth of multi-wavelength data and existing spectroscopy \citep[e.g.,][]{MUSYC,Balestra10} as well as continuous visibility from the MeerKAT site. 

\noindent\underline{Data} The final 5000 hour integration results in a trumpet-shaped data-cube\footnote{The vuvuzela is a uniquely South African trumpet primarily used for encouragement during soccer matches, as most of the world knows after the 2010 World Cup.}, as the primary beam widens with redshift, covering an ever larger volume with redshift.

\begin{figure}
  \begin{center}
    \begin{minipage}[t]{0.5\linewidth}
	\includegraphics[width=\textwidth]{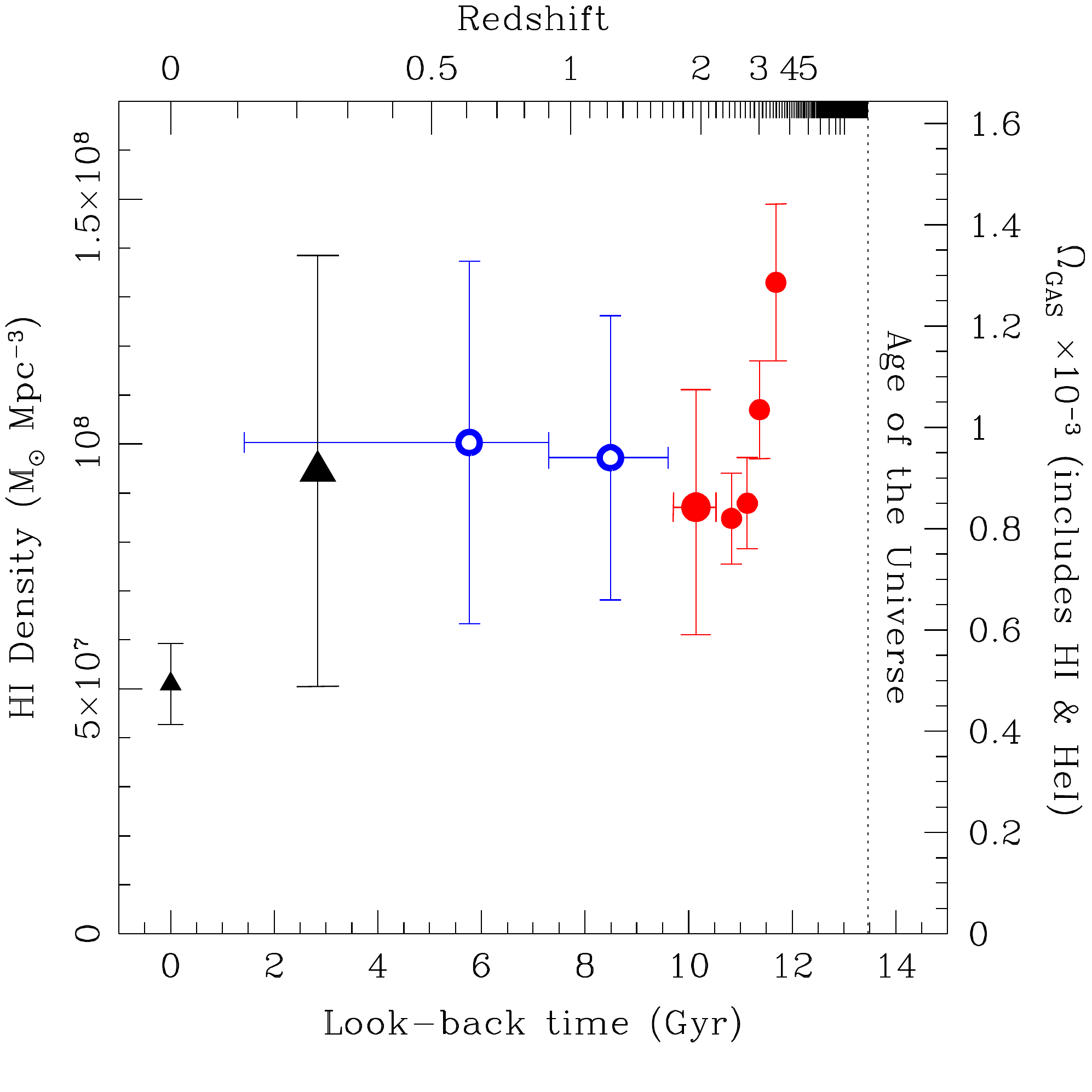}
	\caption{\label{f:Omhi} The cosmic \hi \ density from a variety of methods; direct \hi \ line detections \citep[black tringles][]{Zwaan05, Lah07}, and Ly-$\alpha$ absorption estimates \citep[blue and red circles][respectively]{Rao06, Prochaska05, Noterdaeme09}. }
     \end{minipage}\hfill
     \begin{minipage}[t]{0.49\linewidth}
	\includegraphics[width=\textwidth]{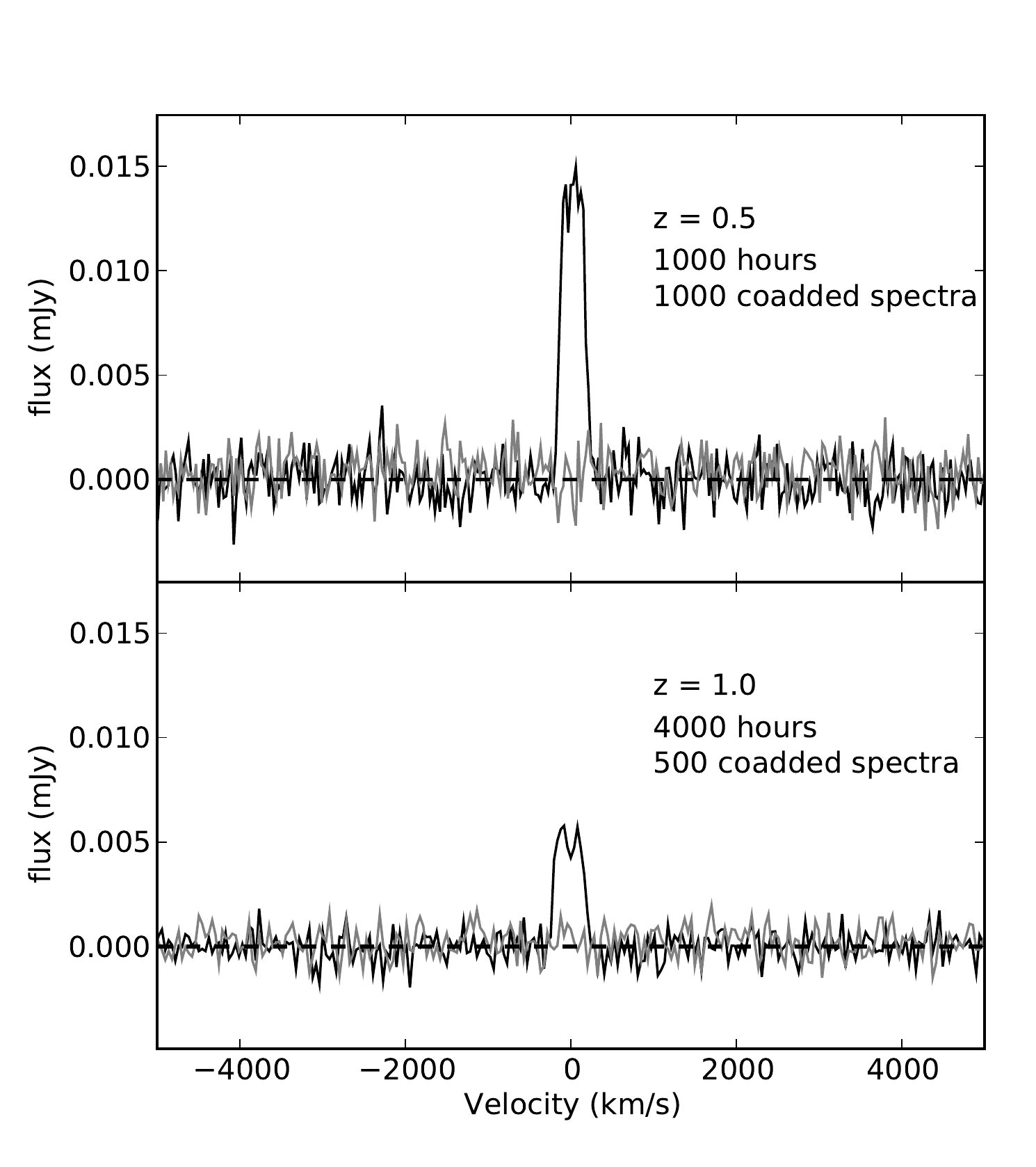}
	\caption{	\label{f:stack} Simulated stacking results (black lines) for z=0.5 (top panel) and z=1.0 (bottom panel), ($\Delta z =0.1$). Galaxies were simulated according to the Oxford $S^{3}$ database \citep{Obreschkow09f}. }
    \end{minipage}
 \end{center}
\end{figure}
%

\noindent\underline{Direct Detections \& Line Stacking} 
Given the shape of the survey volume and the gradually decreasing sensitivity with distance, our science will be done with a mix of direct detections (for low redshifts and/or high \hi \ masses) and the stacking of multiple objects with low signal-to-noise (see also Meyer et al. {\em this volume}). Fig. \ref{f:stack} illustrates the line stacking technique. Using the known positions and redshifts of a sample of sources, their radio spectra are aligned to a common (rest) frequency frame and combined. One can determine the mean \hi \ mass and possibly the \hi \ profile of a sub-class of galaxies. This technique will require an order of magnitude more spectroscopic redshifts than currently available  \citep{MUSYC,Balestra10}.


%

\section{LADUMA Science}

\noindent The LADUMA data will address a multitude of science questions;

\noindent\underline{Cosmic Hydrogen Density ($\Omega_{H \sc I}$)} 
Summing over all objects, one obtains an estimate of the {\em total} atomic hydrogen volume density of the Universe at a given epoch.
Given that the star-formation rate density decreases by a factor ten \citep[][{\em this volume}]{Hopkins08} and specific star-formation rate by a factor three \citep{Noeske07a}, over the redshift range $0 < z \lesssim1.4$, one can expect the star-formation fuel to be depleted, but at what rate? Our direct  \hi \ observations will be compared to those Ly$\alpha$ and Mg II absorbers \citep{Menard09,Kanekar09}.

\noindent\underline{HI Mass Function (\himf)} is the number density of galaxies as a function of their neutral gas content \citep{Zwaan05}. 
An accurate measurement of this quantity is an important parameter in models of galaxy evolution where processes like gas inflow and star formation are sought to be understood. Evolution in the breakpoint, slope, and normalization of the \himf \  can be used to test hierarchical galaxy formation and cold flow models \citep{van-der-Heyden09}.
And finally, we can also explore environmental effects on the \himf \ \citep{Bouchard09a} .

\noindent\underline{Galaxy Evolution}
The \hi \ content of different populations as a function of redshift will place a useful constraint on our understanding of their evolution. We will address questions such as how cold gas mass depends on halo mass? what is the relation between stellar and gas mass over time\citep[e.g.,][]{Kannappan04}, and what is the relation between the cold gas mass and specific star-formation rate? 

\noindent\underline{The Tully-Fisher Relation}; Rotationally supported disks lie on the Tully-Fisher relation \citep{Tully77}. For small high-redshift samples, the stellar T-F relation has been determined \citep[e.g.][]{Kassin07, Puech10}, but the LADUMA data will provide \hi \ data for the first time at higher redshift with the advantage of probing the whole disk (and consequently halo), and with sufficient statistics to determine the normalization and slope of the T-F relation at different redshifts. 


\noindent\underline{Bonus Science} includes the number densities and strengths of OH masers over cosmic time \citep[see][]{Briggs98}, the merger rate from close companions \citep[e.g.,][]{Holwerda11c}, and the possible detection of faint 21 cm absorption (e.g., the ``Cosmic Web''). 

\noindent{\bf Concluding Remarks:}
The LADUMA survey promises to facilitate a range of exciting science and provide the deepest \hi \ observations available until the SKA is commissioned.


\begin{multicols}{2}
\raggedcolumns
\addtocounter{unbalance}{1}

\end{multicols}
\end{document}